# Image Speckle Noise Denoising by a Multi-Layer Fusion Enhancement Method based on Block Matching and 3D Filtering

Huang Shuo[a, b], Zhou Ping[a], Shi Hao[a], Sun Yu[a, c]* and Wan Suiren[a]*

[a] *International Laboratory for Children's Medical Imaging Research, School of Biological Sciences and Medical Engineering, Southeast University, Nanjing, Jiangsu, 210096 China;*

[b] *Shanghai United-imaging Healthcare Co., Ltd, Jiading district, Shanghai, 201807 China;*

[c] *Institute of Cancer and Genomic Science, University of Birmingham, Birmingham, B15 2TT, United Kingdom*

*. Correspondence e-mail: Wan: srwan@seu.edu.cn, sunyu@seu.edu.cn

# Image Speckle Noise Denoising by a Multi-Layer Fusion Enhancement Method based on Block Matching and 3D Filtering

**Abstract:** In order to improve speckle noise denoising of block matching 3d filtering (BM3D) method, an image frequency-domain multi-layer fusion enhancement method (MLFE-BM3D) based on nonsubsampled contourlet transform (NSCT) has been proposed. The method designs a NSCT hard threshold denoising enhancement to preprocess the image, then uses fusion enhancement in NSCT domain to fuse the preliminary estimation results of images before and after the NSCT hard threshold denoising, finally, BM3D denoising is carried out with the fused image to obtain the final denoising result. Experiments on natural images and medical ultrasound images show that MLFE-BM3D method can achieve better visual effects than BM3D method, the peak signal to noise ratio (PSNR) of the denoised image is increased by 0.5dB. The MLFE-BM3D method can improve the denoising effect of speckle noise in the texture region, and still maintain a good denoising effect in the smooth region of the image.

## 1. Introduction

Speckle noise is a common type of noise, which is the main noise in medical ultrasonic images[1-3]. Speckle noise is usually considered as multiplicative noise. The noise intensity in a certain region of image is related to the grayscale value of the noise-free image in that region. Meanwhile, the noise intensity is also related to the texture intensity of that region[4, 5]. In 1980, J. S. Lee proposed the linear approximation form of the fully developed speckle noise, indicating that the fully developed speckle noise can be regarded as the zero-mean additive noise[6].

The denoising technique has been widely studied in a long time, and a lot of efficient approaches has been put forward[7-12]. That can not only improve the visual experience, but also ameliorate the performance of the post-processing algorithms like the segmentation, classification or registration of images. Especially in the medical image analysis, removing the noise in the image helps to acquire more accurate diagnoses [13-14].

One way to image denoising is to improve the property of imaging system and image reconstruction algorithm. For example, optical illumination in the near-infrared optical range is used in the laser photoacoustic imaging system (LOIS-64) to maximize light penetration depth in breast tissue while providing a high absorption contrast of the tumor [15]. E. k. Tan studied the feasibility of using a compact 3T NMR scanner to reduce acoustic noise in the acquisition of diffusion-weighted echo-planar imaging [16].

Moreover, various natural image denoising methods based on this idea have been studied. For example, Y. Geng et al. proposed an optical scanning imaging system based on rotation of a single cylindrical lens, which is an effective method to reconstruct the amplitude and phase information of samples in the axial multi-image computing imaging[17]. In addition, the compressed reflection antenna imaging system proposed by W. Zhang et al. and the light field video generation algorithm proposed by T. C. Wang et al. can also obtain images or videos with low noise[18-19].

The post-processing algorithm after image reconstruction is also an important research field. Wavelet transform denoising methods are effective in image denoising[20-23]. These methods assume that the true (noise-free) signal can be well approximated by a linear combination of few base functions, this means that the signal can be sparsely represented in the transform domain. The real signal energy is concentrated in low frequency band. The noise energy mainly concentrated in high frequency band. Therefore, the noise can be resolved by suppressing high frequency coefficient and the true signal can be effectively estimate. However, wavelet denoising has three main disadvantages: lack of translation invariance, lack of symmetry and poor directional selectivity [24]. Researchers have proposed many improved wavelet denoising algorithms, which can effectively suppress speckle noise in medical images, such as the Fourier-wavelet regularized de-convolution (ForWaRD) method[25] and the robust de-convolution method using higher-order spectral analysis and wavelet[26].

The contourlet transform is a remarkable improvement of wavelet transform, which has the advantages of directivity, anisotropy and low redundancy, and has been widely used in medical image denoising algorithms. The geometric structure of the image can be well represented by the pyramidal directional filter bank (PDFB) with contourlet transform. In PDFB, the images are first decomposed by Laplacian pyramid filter to capture the point-like discontinuities, and then the point-like discontinuities are linked into linear structures using a 2-dimensional directional filter bank (DFB) [27, 28]. However, due to subsampling and upsampling operations in PDFB transform and inverse transform, the contourlet transform is also not translation invariance. Therefore, when used in image denoising, pseudo-Gibbs phenomenon exists due to spectrum aliasing. Based on contourlet transform, in 2005, A. L. Cunha et al. proposed the nonsubsampled contourlet transform (NSCT), which has all the advantages of contourlet transform. At the same time, due to the removal of sampling rate variation in the NSCT filter bank, it also has the translation invariance. Similar to the contourlet transform, NSCT can also be divided into two parts: a nonsubsampling pyramid structure to realize multiscale resolution and a nonsubsampling DFB structure to provide directionality[29, 30]. Figure 1 shows the NSCT nonsubsampling filter bank (NSFB) structure and its idealized frequency partitioning.

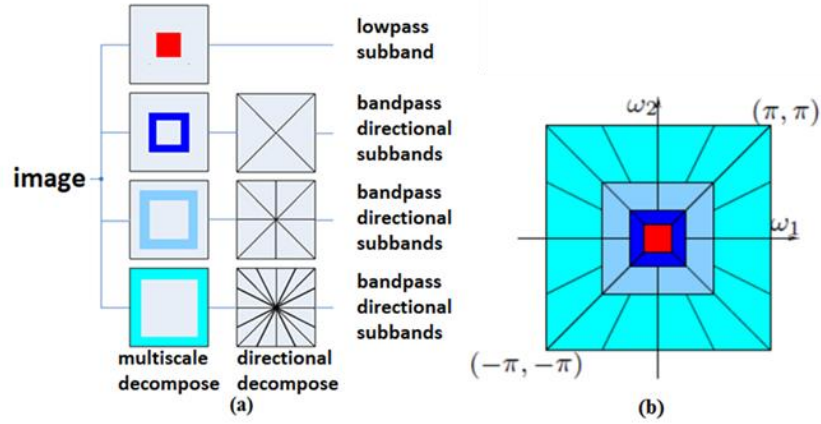

**Figure 1** Nonsubsampled contourlet transform. (a) NSFB structure that implements the NSCT. (b) Idealized frequency partitioning obtained with the NSFB structure. Figures 1(a) and 1(b) are from Reference [29].

The Bayes least squares-Gaussian scale mixtures model (BLS-GSM) wavelet denoising method is another efficient wavelet transform-domain denoising method[31], which is based on a statistical model of the coefficients of an overcomplete multiscale oriented wavelet basis. Its specific denoising steps are: (1) decompose the image into pyramid subbands at different scales and orientations; (2) denoise each subband, except for the lowpass residual band; and (3) invert the pyramid transform, obtaining the denoised image.

In recent years, non-local algorithms for image denoising have been widely studied since they can obtain excellent image restoration effects[32-35]. One famous nonlocal algorithm is the nonlocal means (NL-means) method[32], which removes the zero mean additive gaussian white noise of a given pixel in an image by calculating the weighted average gray value of all pixels in the image. The weight of each pixel is determined by the similarity between the grayscale intensity vector of the pixels in a square window centered by it and that of pixels in the squire window with the same size centered by the pixel to be denoised. Since fully developed speckle noise can be regarded as zero-mean additive noise, these methods have good performance in removing fully developed speckle noise.

The block matching and 3D filtering (BM3D) method [20, 36] combines the advantages of the non-local denoising algorithm with that of the denoising method in transformation domain, therefore, it can obtain better denoising results than other non-local denoising methods and transform-domain denoising approaches [10]. In order to improve the performance of BM3D method, some improved methods have been proposed [37-42]. However, although BM3D method has better performance in suppressing speckle noise, due to the over-smoothed distribution of the speckle noise, some high-frequency information such as texture and edge will be over-smoothed when denoising images through BM3D method, resulting in loss of the image information.

In order to improve speckle noise denoising performance of block matching and 3d filtering (BM3D) method, an image frequency-domain multi-layer fusion enhancement method (MLFE-BM3D) based on nonsubsampled contourlet transform (NSCT) has been proposed. The method designs a NSCT hard threshold denoising enhancement to preprocess the image, then uses fusion enhancement in NSCT domain to fuse the preliminary estimation results of images before and after the NSCT hard threshold denoising, finally, BM3D denoising is carried out with the fused image to obtain the final denoising result. Experiments on natural images and medical ultrasound images show that MLFE-BM3D method can achieve better visual effects than BM3D method, the peak signal to noise ratio (PSNR) of the denoised image is increased by 0.5dB.The MLFE-BM3D method can improve the denoising effect of speckle noise in the texture region, and still maintain a good denoising effect in the smooth region of the image.

## 2. Method

The BM3D method contains two main steps [20, 36], which are basic estimation (step 1 in Figure 2) and final estimation (step 2 in Figure 2). The detailed steps are similar in each main step, which are grouping of the 2D blocks (images in each group form a 3D array), calculating local estimates by the three-dimensional collaborative filtering in the spectrum domain (3D hard-threshold method in step1, and Wiener filtering in step 2) of each 3D array and the aggregation of the weighted means of the local estimates. The preliminarily denoised basic estimate results are employed in the step 2 to improve the accuracy of grouping and to acquire more accurate filtering results by using it as the pilot signal of the empirical Wiener filtering. Figure 2 shows the flowchart of BM3D algorithm.

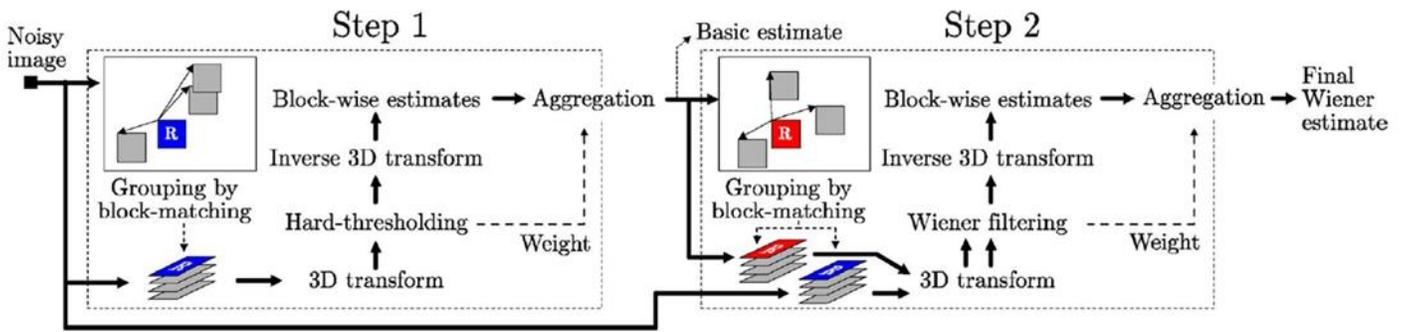

**Figure 2** The flowchart of the BM3D algorithm. Figure 2 is from Reference [20].

An improved BM3D method for speckle noise with the image multi-layer fusion and enhancement has been proposed in this work, which is named as "the MLFE-BM3D method". The following three steps are employed in the MLFE-BM3D method to improve BM3D method. Firstly, the NSCT hard threshold denoising and enhancement is used to preprocess the original noised image and improve the matching accuracy in the estimation. Secondly, the basic estimation results of images before and after the NSCT hard threshold denoising and enhancement are decomposed by NSCT, and the image multi-layer fusion based on the geometric mean is implemented to merge the corresponding layers to help protect the texture and boundaries while suppressing the noise in the smooth regions. The third is to amplify the coefficients on mid-frequency layers of the fused layers to enhance the details of the image and increase the matching accuracy of the texture areas. The block diagram of the MLFE-BM3D method is shown in Figure 3.

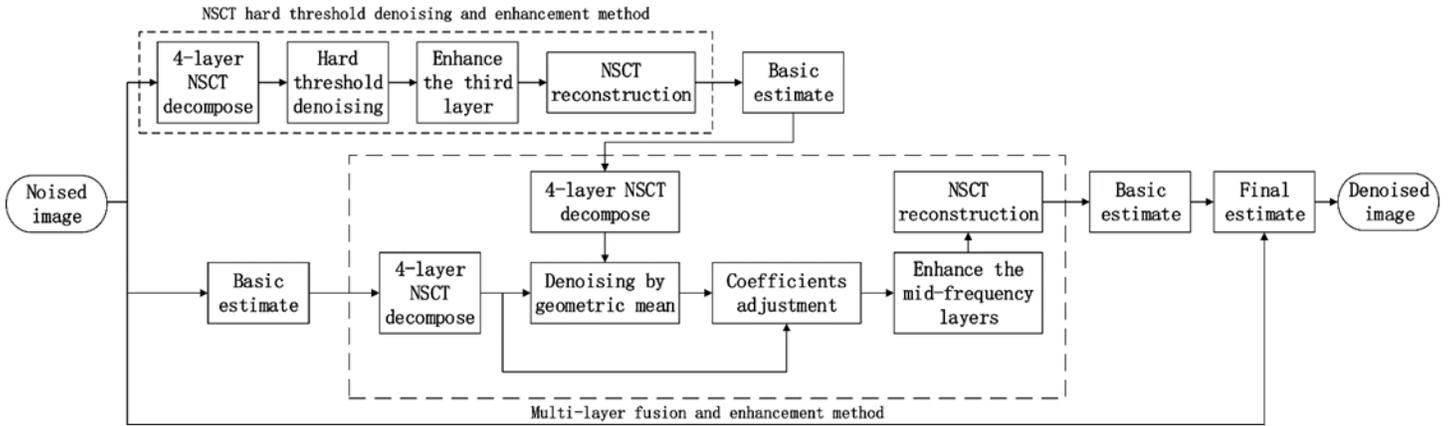

**Figure 3** The block diagram of MLFE-BM3D method.

**Table I** The Specific Process of the NSCT Hard Threshold Denoising and Enhancement Method

| **Algorithm 1**. NSCT hard threshold denoising and enhancement |
|---|
| 1. Input: The noised image $u_0$. |
| 2. Decompose $u_0$ by 4-layer NSCT transform. |
| 3. for $i = 1:1:4$ |
| 4.    NSCT hard threshold denoising: $Y(c_i) = \text{HardThreshold}(c_i)$. |
| 5. End |
| 6. $Y(c_3) = Y(c_3) \times 2$ |
| 7. NSCT reconstruction. |
| 8. Output: The reconstructed image $u_R$. |

The specific process of the NSCT hard threshold denoising and enhancement method is shown in Table I. The number of directions in each NSCT decomposed layer is 1, and $c_i$ denotes the $i$-th layer. In this step, firstly, we denoise the NSCT decomposed coefficients by the $K$-sigma thresholding hard threshold denoising method [29], as shown in Equation (1):

$$Y(c_i(x,y)) = \text{HardThreshold}(c_i(x,y)) = \begin{cases} c_i(x,y), & |c_i(x,y)| \geq K\sigma_{c_i} \\ 0, & |c_i(x,y)| < K\sigma_{c_i} \end{cases} \quad (1)$$

where $\sigma_{c_i}$ denotes the standard deviation of the noise in the $i$-th layer, which can be estimated through: $\sigma_{c_i} = \text{Median}(|c_i|)/0.6745$, where $\text{Median}(|c_i|)$ denotes the median value of all absolute coefficients in the $i$-th layer. $K = 4$ in the first layer, and $K = 3$ in the rest layers [43]. Then we enlarge the coefficients of the third layer in the 4-layer NSCT decomposition to increase the intensity of the texture and boundaries of the image:

$$Y(c_3) = Y(c_3) \times 2 \quad (2)$$

This step can improve the accuracy of the block matching and better protect the texture and boundaries in the estimation result of the NSCT denoised image. This is because that in the decomposed layers, the 3rd layer contains coefficients of the high-frequency textures and boundaries, enhance this layer can improve the accuracy of grouping in the basic estimate, thereby improve the denoising performance of the estimate. And although the forth layer contains coefficients of higher-frequency textures, the energy of noise is also mostly concentrated in this layer, so that the 4-*th* layer is not enhanced. However, some noises remain inevitably in the image after NSCT hard threshold denoising. These noises will affect the matching accuracy of the smooth region, which results in a reduction of the denoising performance of the smooth region. The enhancement operation will amplify these noises, which will further reduce the denoising performance of the smooth areas.

Table II The Specific Process of the Multi-layer Fusion and Enhancement Method

| **Algorithm 2**. The multi-layer fusion and enhancement method |
| --- |
| 1. Input: The basic estimation result $u_{on}$ of noised image. |
| 2. Input: The basic estimation result $u_{oR}$ of image $u_R$. |
| 3. Decompose $u_{on}$ by 4-layer NSCT transform. |
| 4. Decompose $u_{oR}$ by 4-layer NSCT transform. |
| 5. for $i = 1 : 1 : 4$ |
| 6.     Denoising by calculating the geometric mean $c_{fi}$ by Equation (3). |
| 7.     Adjust the coefficients in each layer by Equation (4). |
| 8.     Enhance the mid-frequency layers by Equation (5). |
| 9. end |
| 10. NSCT reconstruction. |
| 11. Output: The reconstructed image $u_F$. |

To solve this problem, we designed a multi-layer fusion and enhancement method, which is shown in Table II. In this method, a four-layer NSCT decomposition has been implemented on the basic estimation results of the original noised image $u_{on}$ and the image $u_{oR}$ after the NSCT hard threshold denoising and enhancement, respectively. The number of directions in each NSCT decomposed layer is also 1. The decomposition results are denoted as $c_{n1}, \ldots, c_{n4}$ for $u_{on}$ and $c_{R1}, \ldots, c_{R4}$ for $u_{oR}$, respectively. The geometric mean $c_{fi}$ of the coefficients at each corresponding position on these two groups of decomposed layers is calculated through Equation (3) to eliminate the residual noise in the smooth region, where $i$ is the number of layers, $k$ and $j$ represent the length and width in pixel of each layer, sgn() denotes the sign function.

$$c_{fi} = \text{sgn}(c_{ni}(k,j)) \times \sqrt{|c_{ni}(k,j) \times c_{Ri}(k,j)|}, \ i = 1,2,3,4 \tag{3}$$

Since the calculation of the geometric mean both reduces the coefficients of noise and useful signal, in order to reduce the loss of the useful signal, Equation (4) is employed to adjust the coefficients in each layer of $c_{fi}$ using the range of the coefficients in the corresponding layer of $c_{ni}$.

$$c'_{Fi} = \frac{c_{fi} - \min(c_{fi})}{\max(c_{fi}) - \min(c_{fi})} \times (max(c_{ni}) - min(c_{ni})) + min(c_{ni}) \qquad (4)$$

The denoising performance of the basic estimate is closely relevant to the accuracy of grouping in BM3D. The noise may obscure the image and influence the grouping accuracy, especially in the boundaries or textured regions. So, in order to enhance these regions and acquire better grouping results, the coefficients of the second and third layers are enhanced, as shown in Equation (5).

$$c_{Fi} = \begin{cases} c'_{Fi}, & i = 1,4 \\ 2 \times c'_{Fi}, & i = 2,3 \end{cases} \qquad (5)$$

Since the first layer is the low-frequency layer, and the energy of noise is concentrated in the fourth layer, these two layers are not enhanced. And on comparison with the noised image, the data after the coefficient adjustment contain less noise, so more layers are enhanced than the NSCT hard threshold denoising and enhancement method. The reconstructed image $u_F$ of the amplified coefficients $c_{Fi}$ is obtained next, which is the result of the multi-layer fusion and enhancement method.

Using the multi-layer fusion and enhancement method, the coefficients of the high-frequency regions are enhanced, and the residual noise in the smooth regions is suppressed, which improves the matching accuracy in the following basic estimation and final estimation. Meanwhile, using the image $u_F$ whose high-frequency regions are amplified as the Wiener filter's pilot signal in the final estimation can better protect the texture areas and boundaries, thereby further improving the image's denoising effect.

3. Results

In order to verify the denoising performance of MLFE-BM3D method, the denoising effects of MLFE-BM3D, BM3D and BLS-GSM on Lena image of size 512 × 512 with speckle noise $\sigma^2 = 1300$ have been obtained and compared. In the BM3D and MLFE-BM3D methods, when transforming the matched three-dimensional matrix into the three-dimensional transformation domain in the "basic estimation" and "final estimation", the adopted transforms are bior1.5 wavelet transform and discrete cosine transform, respectively.

In order to compare the performance of each method quantitively, the signal-to-noise ratio (SNR), the peak signal-to-noise ratio (PSNR), the root mean square error (RMSE), and the mean structural similarity index measurement (MSSIM) [44-48] are employed. These parameters can describe the similarity between the original clean and the denoised images from different aspects. These parameters are calculated via Equations (6-9), where $u_t$ and $u$ are the denoised and the original clean images respectively, $M$ is the number of pixels in each image, $\mu$ and $\sigma$ are the mean value and standard variation of image respectively, $cov(u_{t,i}, u_i)$ denotes the covariance of images in square windows centered by pixel $i$ in images $u_t$ and $u$, max is the maximum possible pixel value of the image. In our experiments, the pixels are represented using 8 bits per sample, so max = 255. $c_1$, $c_2$ and $c_3$ are three constant parameters set, and $c_1 = (0.01 \times max)^2$, $c_2 = (0.03 \times max)^2$, $c_3 = 1/2 \times c_2$.

$$SNR(u_t, u) = 10 \times \log_{10}\left(\frac{\sum_{i=1}^{M}(u_t(i)-\mu(u_t))^2}{\sum_{i=1}^{M}(u_t(i)-u(i))^2}\right) \tag{6}$$

$$PSNR(u_t, u) = 10 \times \log_{10}\left(\frac{max^2}{\sum_{i=1}^{M}(u_t(i)-u(i))^2}\right) \tag{7}$$

$$\text{RMSE}(u_t, u) = \sqrt{\frac{\sum_{i=1}^{M}(u_t(i)-u(i))^2}{M}} \tag{8}$$

$$\text{SSIM}(u_{t,i}, u_i) = \frac{2\mu(u_{t,i})\mu(u_i)+c_1}{\mu^2(u_{t,i})+\mu^2(u_i)+c_1} \cdot \frac{2\sigma(u_{t,i})\sigma(u_i)+c_2}{\sigma^2(u_{t,i})+\sigma^2(u_i)+c_2} \cdot \frac{cov(u_{t,i},u_i)+c_3}{\sigma(u_{t,i})\sigma(u_i)+c_3} \tag{9a}$$

$$\text{MSSIM}(u_t, u) = \frac{1}{M}\sum_{i=1}^{M}\left(\text{SSIM}(u_{t,i}, u_i)\right)^2 \tag{9b}$$

The difference on SSIM index maps of results of BM3D and MLFE-BM3D are calculated to acquire detailed information about the denoising improvement on regions with different smoothness. The SSIM index map is a matrix that has the same size with the image [45, 48]. To calculate the value of its elements, a local 8×8 square window which moves pixel-by-pixel over the entire image is used. At each step, the local statistics and SSIM index are calculated within the local window by Equation 9(a). And after the calculation of the SSIM index map, the mean value of the SSIM index map is calculated by Equation 9(b) to evaluate the overall image quality, which is the value of parameter MSSIM.

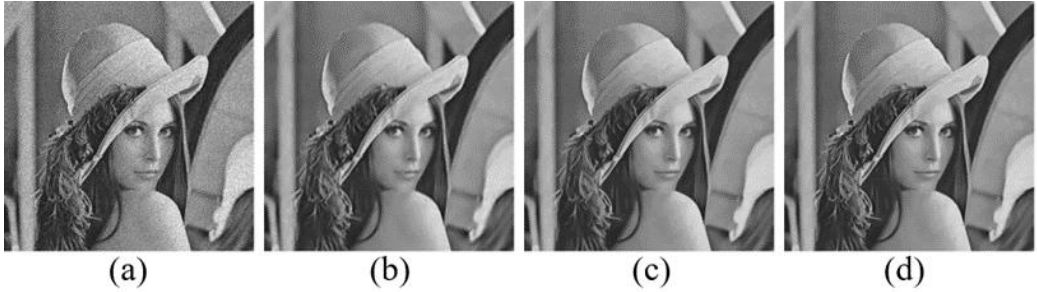

**Figure 4** The denoising effects of different methods. (a) is the noised image; (b)-(d) are the denoising results of BLS-GSM method, BM3D method and MLFE-BM3D method, respectively.

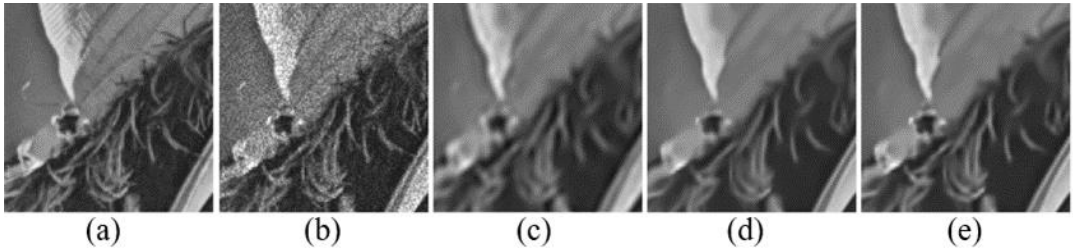

**Figure 5** The local image of the denoising image. (a) is the clean image; (b) is the noised image; (c)-(e) are the local image of the denoised image by BLS-GSM method, BM3D method and MLFE-BM3D method, respectively.

As shown in Figure 4, all the three approaches can achieve acceptable noise removal results in the smooth regions. However, as for the textured zones, the MLFE-BM3D method can better protect the texture information.

In order to make a better comparison, the local images of Figure 4(a)-(d) have been enlarged and shown in Figure 5. Figure 5 shows that the texture regions are smoothed in the results of all the three methods. However, the MLFE-BM3D method can retain more texture information. Therefore, the MLFE-BM3D method has better denoising performance of the denoising of texture regions.

Table III shows the quantitative evaluation results of the denoising performances of different methods. Parameters of the image denoised by the MLFE-BM3D method are superior to those of the other two methods, which shows that MLFE-BM3D method can achieve better denoising results. Moreover, on comparison of the denoising performance of the whole image and the textured local image, the improvement of the MLFE-BM3D method is more significant in the textured local image.

However, the denoising performances of all these methods in the textured local image are worse than their performances in the whole image, which indicates that the denoising performances of these methods need to be further improved, which is our future work.

**Table III** The denoising performances of different approaches.

| Method | SNR (dB) | PSNR (dB) | RMSE | MSSIM |
|---|---|---|---|---|
| Lena whole image, $\sigma^2 = 1300$ | | | | |
| Noised image | 17.0294 | 22.6857 | 18.7178 | 0.4233 |
| BLS-GSM | 24.7610 | 30.4173 | 7.6855 | 0.8339 |
| BM3D | 26.4171 | 32.0735 | 6.3514 | 0.8587 |
| MLFE-BM3D | 26.9319 | 32.5882 | 5.9859 | 0.8628 |
| Lena local image, $\sigma^2 = 1300$ | | | | |
| Noised image | 17.0032 | 24.4665 | 15.2481 | 0.6590 |
| BLS-GSM | 19.0006 | 26.4638 | 12.1157 | 0.7255 |
| BM3D | 20.5193 | 27.9826 | 10.1722 | 0.7732 |
| MLFE-BM3D | 21.4501 | 28.9134 | 9.1384 | 0.7938 |

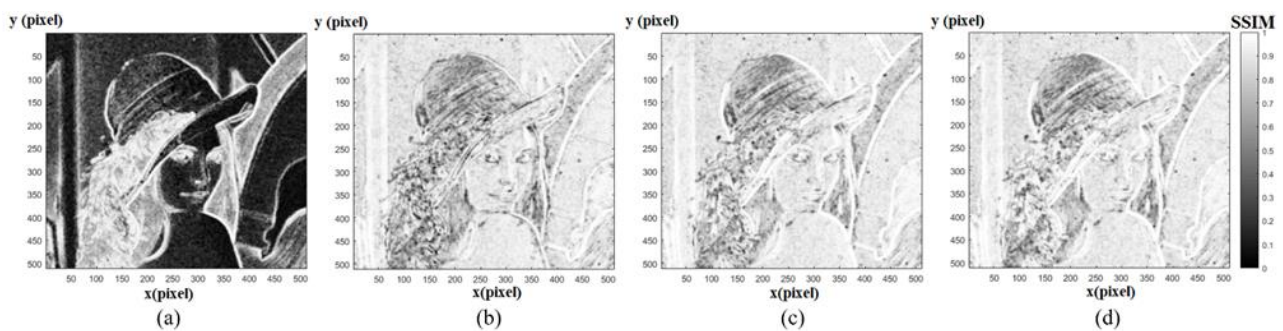

**Figure 6** The SSIM index maps of denoising results of different methods in Figure 4. (a) is the SSIM index map noised image; (b)-(d) are the SSIM index maps of denoising results of BLS-GSM method, BM3D method and MLFE-BM3D method, respectively.

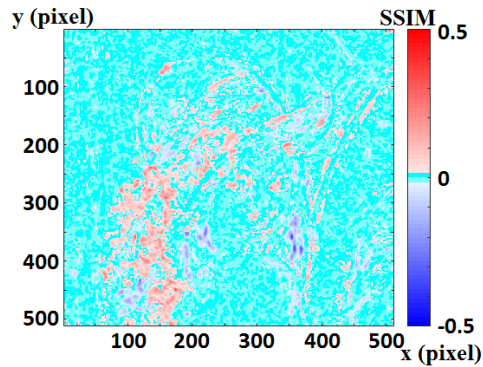

**Figure 7** The differences in SSIM index map between results of the BM3D method and MLFE-BM3D method in Figure 4(c) and (d). In this figure, the red regions denote the regions that MLFE-BM3D method has better performance, while in the blue regions, the BM3D method is better.

The SSIM index maps of the denoising results in Figure 4 are shown in Figure 6. The result of MLFE-BM3D method has similar SSIM indexes in the smooth regions, and better indexes in the textured regions than the BLS-GSM method and the BM3D method. To better compare the maps of BM3D and MLFE-BM3D methods, the SSIM index map of the result of MLFE-BM3D method (shown in Figure 6(c)) is subtracted by that of the result of BM3D method (shown in Figure 6(d)) to test the improvement of MLFE-BM3D method on regions with different smoothness, and results are shown in Figure 7, where the red regions are the regions that MLFE-BM3D method has better performance, while in the blue regions, the BM3D method is better. Figure 7 shows that the MLFE-BM3D method can obtain better denoising results in the textured zones. And as for the smooth regions, the MLFE-BM3D method has similar denoising performance as BM3D method. Only in a few regions the BM3D method overcomes the MLFE-BM3D method.

## 4. Discussion

In this section, the natural images and the medical ultrasonic images of liver and muscle layer of arm are used to test the denoising performance of the MLFE-BM3D method. The denoising performances of MLFE-BM3D, BM3D and BLS-GSM methods are compared through quantitative evaluation parameters and the visual performances.

### 4.1 Compare with natural images

Tests on Lena and Baboon images with speckle noise whose variance ranges from 300 to 3200 have been done to verify the MLFE-BM3D method. The SNR, PSNR, RMSE and MSSIM are used to quantitatively evaluate the results, and the line charts about the performance changes of the BLS-GSM, BM3D and MLFE-BM3D with different variance of noise are shown in Figures 9 to 12. The original clean image and the local image of Baboon are shown in Figure 8, Figure 5(a) is the local image of Lena.

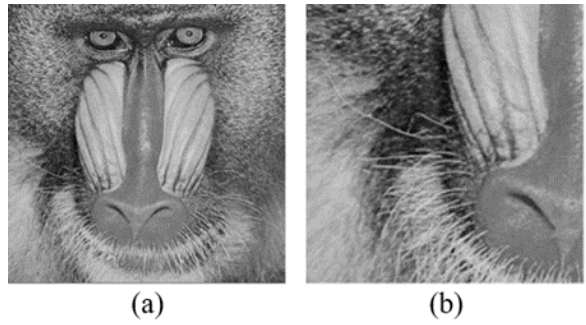

**Figure 8** The original clean image and the local image of Baboon. (a) is the original clean image, (b) is the local image.

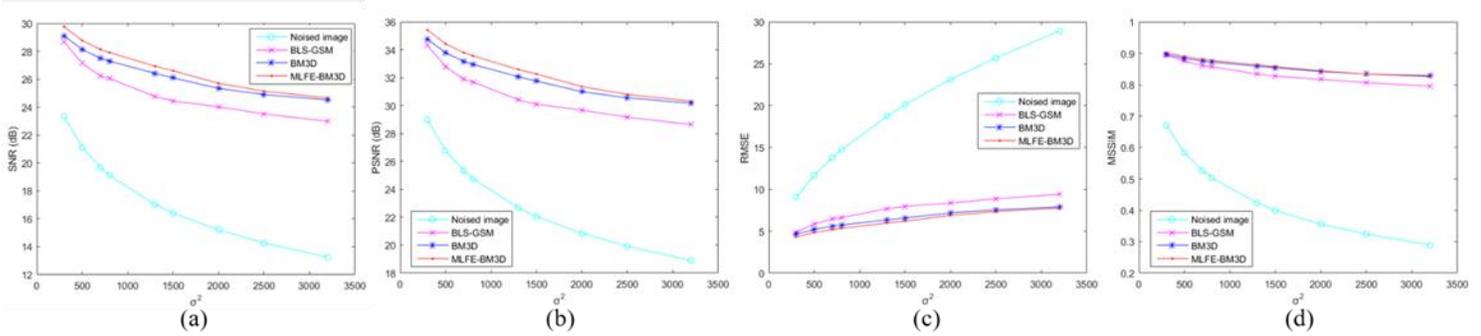

**Figure 9** The performance of BLS-GSM, BM3D and MLFE-BM3D method on the whole image of Lena. (a)-(d) are the line charts of SNR, PSNR, RMSE and MSSIM, respectively.

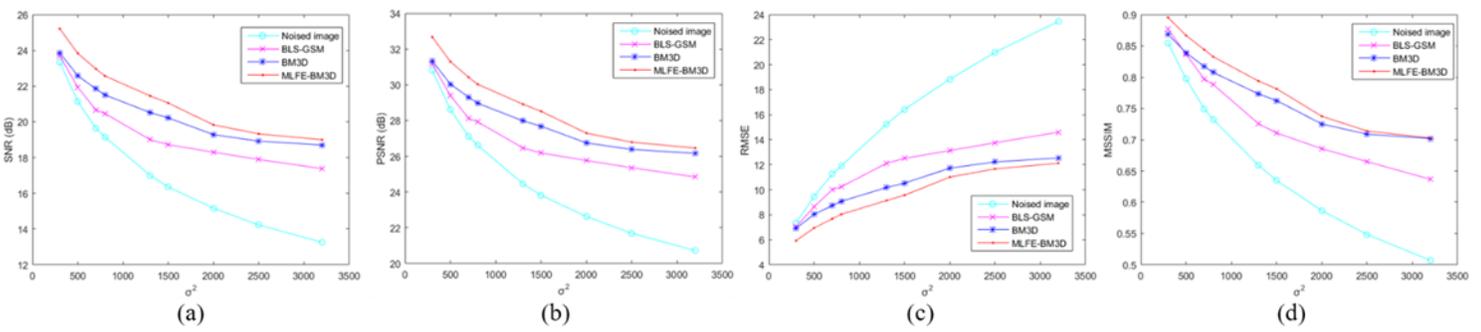

**Figure 10** The performance of BLS-GSM, BM3D and MLFE-BM3D method on the local image of Lena. (a)-(d) are the line charts of SNR, PSNR, RMSE and MSSIM, respectively.

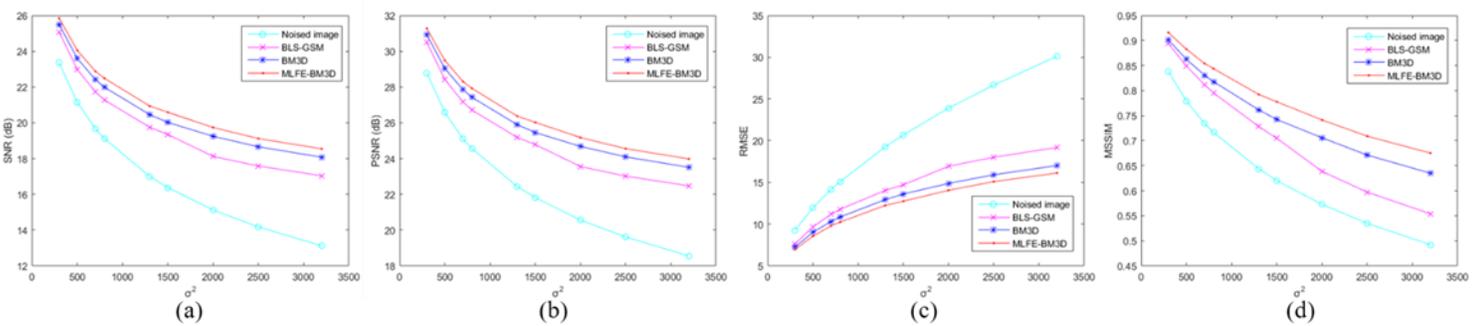

**Figure 11** The performance of BLS-GSM, BM3D and MLFE-BM3D method on the whole image of Baboon. (a)-(d) are the line charts of SNR, PSNR, RMSE and MSSIM, respectively.

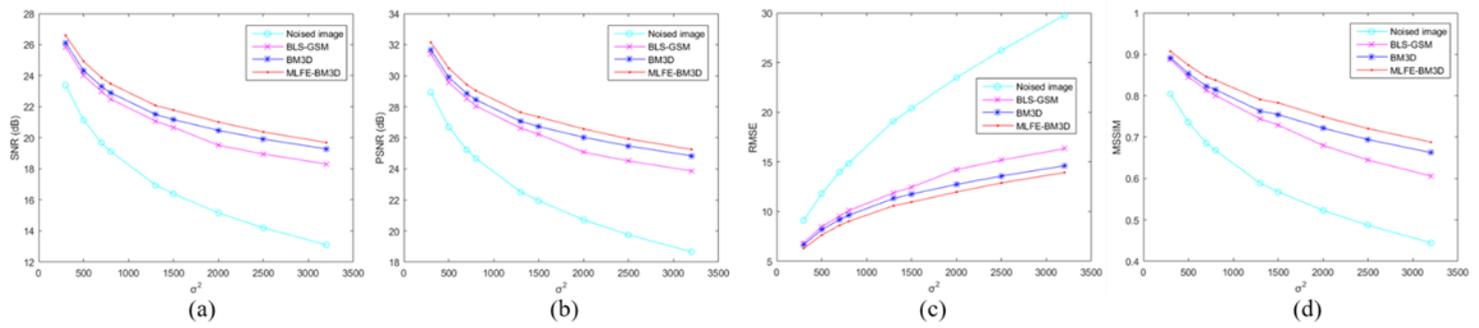

**Figure 12** The performance of BLS-GSM, BM3D and MLFE-BM3D method on the local image of Baboon. (a)-(d) are the line charts of SNR, PSNR, RMSE and MSSIM, respectively.

The results of MLFE-BM3D are better in almost all situations. The only exception is that when $\sigma^2 = 3200$, the BM3D scores a better MSSIM. However, under that condition, other parameters of MLFE-BM3D are still better. Moreover, Figures 9 to 12 show that the improvement on MLFE-BM3D than BM3D in the textured regions of the image is larger than that on the corresponding whole images. This result further indicates that the MLFE-BM3D method has better performance in the denoising of textured regions. Furthermore, the greater improvement on the Baboon image which contains more textured regions than the Lena image also demonstrates that the MLFE-BM3D method is more suitable to the images with more textured regions.

**4.2 Compare with medical ultrasonic images**

The medical ultrasonic data are also used to test the performance of the MLFE-BM3D. The dataset contains four liver images and two images of the muscle layer of arm. The original noised images and denoised images by BM3D and MLFE-BM3D are shown in Figures 13 and 14, respectively. The denoised results show that the boundaries in denoised images is similar between the denoised images by BM3D and MLFE-BM3D. However, the smooth regions in results of MLFE-BM3D are smoother than those by BM3D, which indicates that the MLFE-BM3D method has good performance in the protection of boundaries and textured regions, so that the smooth areas can be better smoothed and denoised with similar denoised results in the textured regions.

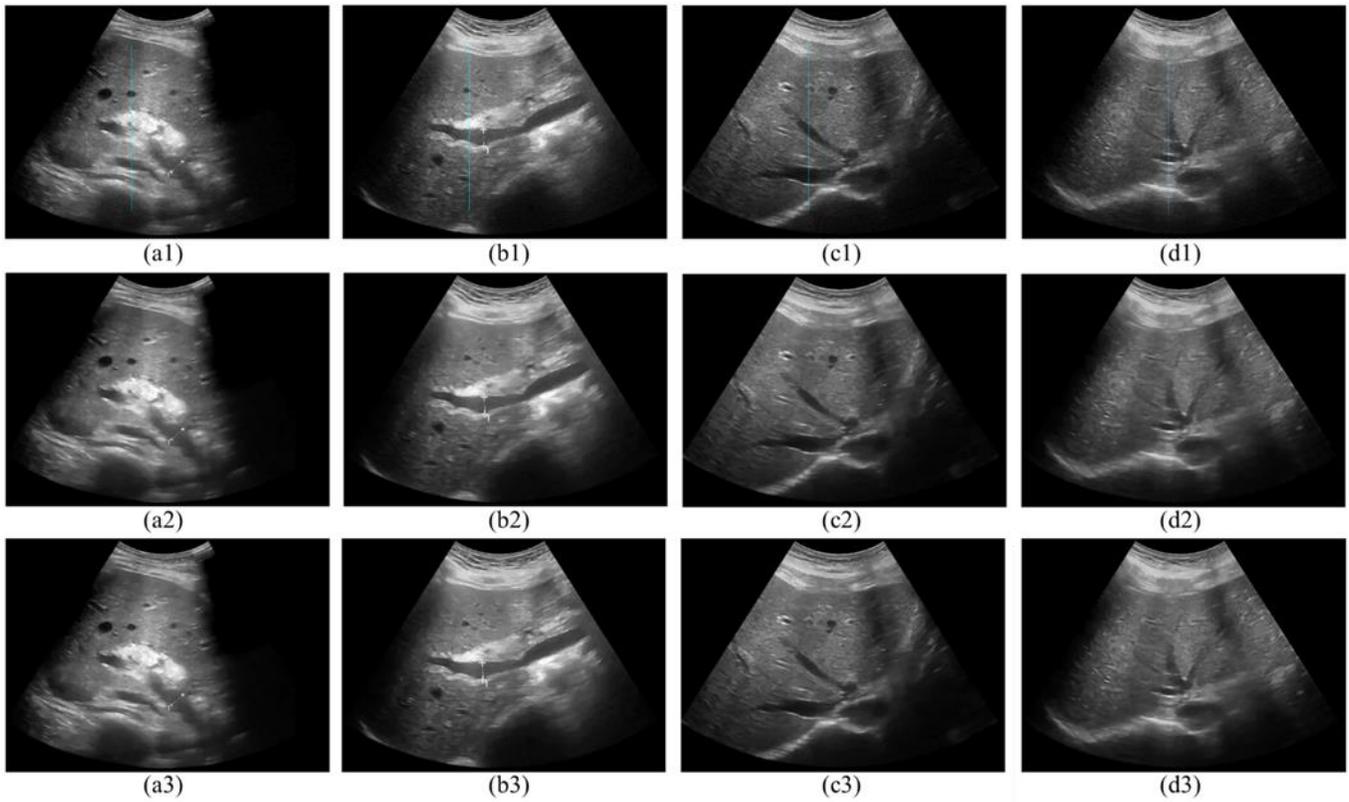

**Figure 13** The performance of BM3D and MLFE-BM3D on medical ultrasonic images of liver. (a1)(b1)(c1)(d1) are the original noised image, (a2)(b2)(c2)(d2) are the denoised results of BM3D method, (a3)(b3)(c3)(d3) are the denoised results of MLFE-BM3D method.

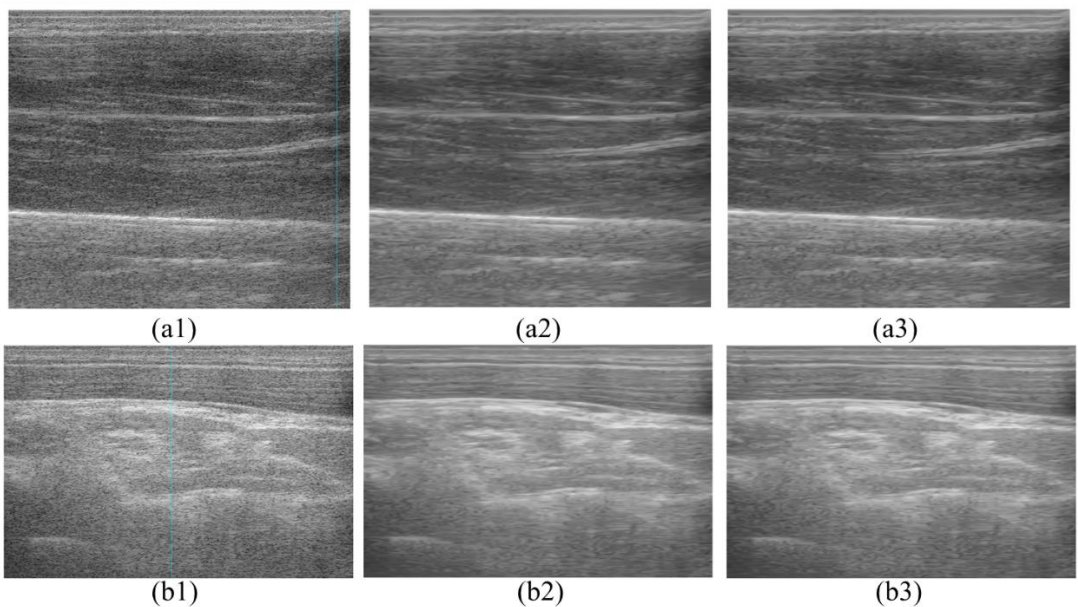

**Figure 14** The performance of BM3D and MLFE-BM3D on medical ultrasonic images of the muscle layer of arm. (a1) (b1) are the original noised image, (a2) (b2) are the denoised results of BM3D method, (a3) (b3) are the denoised results of MLFE-BM3D method.

Furthermore, the marks in the noised image, such as the white imaginary line, the plus sign "+" and the number "1" in Figure 13(a1) and (b1) are still clear in the results of both BM3D method and MLFE-BM3D method, which also indicates that these two methods are good at the protection of textures and details.

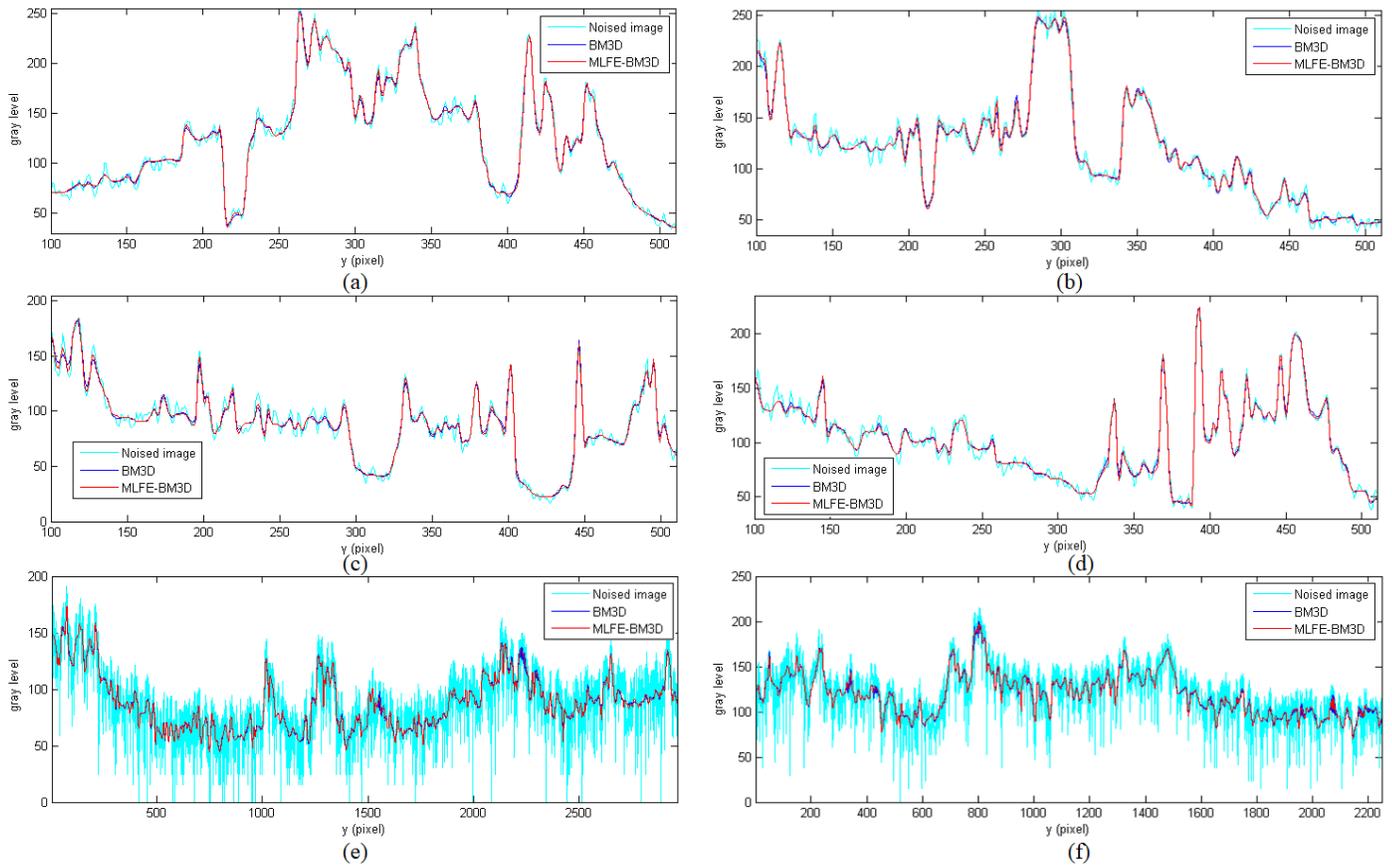

**Figure 15** The gray level distribution curves of results in of pixels on the cyan lines of Figure 13(a1) (b1) (c1) and (d1) as well as Figure 14 (a1) and (b1). (a)-(f) are curves of Figure 13(a1) (b1) (c1) (d1) and Figure 14 (a1) (b1), respectively.

The gray level distribution curves of results in of pixels on the cyan lines of Figure 13(a1) (b1) (c1) and (d1) as well as Figure 14 (a1) and (b1) are shown in Figure 15. On comparison with the results of the BM3D, the curves of MLFE-BM3D have better performance in the suppressing of the striped artefact while the peaks on the curve of MLFE-BM3D have similar amplitude than those on the curve of BM3D. Moreover, when it comes to the muscle layer's images with severe speckle noise and striped artefact, the improvement of the MLFE-BM3D method is significant.

According to the analysis above, MLFE-BM3D method improves the denoising performance of BM3D method on speckle noise, especially when the image has more high-frequency parts. However, the MLFE-BM3D method still has some shortcomings. The major one is that it has long calculation time, which is several seconds, depending on the size of image. Although comparing with the BM3D method, the calculation time of the MLFE-BM3D method only increases slightly for less than 1 second, we still need to work on reducing its calculation time.

## 5. Conclusion

In order to improve the denoising performance of BM3D method on speckle noise and reduce the loss of detailed information, an image frequency-domain multi-layer fusion enhancement method (MLFE-BM3D) based on nonsubsampled contourlet transform (NSCT) has been proposed. The NSCT hard threshold denoising and enhancement method is used to pre-process image, and a multi-layer fusion and enhance method in the NSCT transform domain has been used to improve the grouping accuracy of BM3D. Finally, the BM3D method has been employed to denoise the fused image. Experiments on both the natural images and the medical ultrasonic images show that the MLFE-BM3D method can improve speckle noise denoising performance on the textured regions. Meanwhile, this method maintains a good denoising performance on the smooth regions.

However, this method also has a disadvantage that the time-consuming problem of the BM3D algorithm has not been improved. Therefore, the future work will focus on reducing the computation time. The deep learning algorithms will also be focused to improving the grouping accuracy and the denoising effect [49-51].

## Acknowledgment

The authors thank Prof. Siyuan He, Mr. Yu Yunlei, Mr. Zhang Yi form Southeast University and Mr. Fu Tianyu from Shandong University for their useful discussions and helpful suggestions. We also thank the Beijing Tsinghua Changgung Hospital for providing the medical ultrasonic images of liver. Moreover, the authors thank the anonymous reviewers for their helpful comments and suggestions, thank the engineers and researchers in Alibaba who have given the useful suggestions.

## Funding

This work was supported by the National Key R&D Program of China (grant 2017YFC0112801), the National Natural Science Foundation of China (grants 61771130, 51576115, 61127002, 11572087 and 31300780), the Research Centre for Learning Science of Southeast University (grant 3207038391), the School of Biological Sciences and Medical Engineering of Southeast University (grant 3207037434), the Project of Texas Instruments (TI) Company: Development of Modular OCT Imaging System based on TI Chips (grant 8507030129) and the Key Project of Special Development Foundation of Shanghai Zhangjiang National Innovation Demonstration Zone (grant 1701-JD-D1112-030).

## Conflict of Interest

The authors declare no conflict of interest.